# Chapter 1

# Physics with Trapped Charged Particles

Martina Knoop[1], Niels Madsen[2] and Richard C. Thompson[3]

[1]*CNRSand Université d'Aix-Marseille,
Centre de Saint Jérôme, Case C21,
13397 Marseille Cedex 20, France
Martina.Knoop@univ-amu.fr*

[2]*Department of Physics, College of Science, SwanseaUniversity,
Swansea SA2 8PP, United Kingdom
n.madsen@swansea.ac.uk*

[3]*Department of Physics, Imperial College London,
London SW7 2AZ, United Kingdom
r.thompson@imperial.ac.uk*

Ion traps, which were first introduced in the late 1950s and early 1960s, have established themselves as indispensable tools in many areas of physics, chemistry and technology. This chapter gives a brief survey of the operating principles and development of ion traps, together with a short description of how ions are loaded and detected. This is followed by a brief account of some of the current applications of ion traps.

## 1.1. Introduction

When Wolfgang Paul and Hans Dehmelt were developing the first ion traps in the late 1950s and early 1960s, it is unlikely that they expected to receive Nobel Prizes for this work in 1989. In 2012 another Nobel Prize was awarded in the area of ion traps, to David Wineland. These awards demonstrate how important ion traps have become in the 50 years since their introduction.

The Winter School on "Physics with Trapped Charged Particles" was held at the Les Houches centre in France from 9–20 January 2012. More than 20 speakers gave lectures on a wide variety of topics including the basic principles of ion traps (including storage rings); techniques for cooling, manipulating and detecting the ions;and highly specialized applications such as precision mass measurements and quantum information processing.

Students at the School, including PhD students and postdoctoral researchers, benefited from tutorials with the lecturers as well as the more formal lectures, and were able to have extended discussions with the lecturers outside the timetabled sessions.

This book includes lecture notes from many of the lecturers at the School, though not all were able to contribute. The organizers would like to express their thanks to all the lecturers who have contributed to this review volume, which we hope will be a useful resource for those who attended the School, as well as to newcomers starting out in this exciting field.

This chapter does not attempt to give a thorough review of ion traps; rather, it is to be seen as a short introduction to the subject, giving a flavour of the physics of ion traps and their applications. No attempt is made to give full references, especially as the other chapters in the book give much more detail than can be covered here. For details of the background to ion traps and general techniques that are used for trapped ions, the reader should refer to the books by Werth and collaborators[1,2] and Ghosh.[3]

## 1.2. History of Ion Traps

The two main types of ion trap (the Paul, or radiofrequency, trap and the Penning trap) were both introduced at about the same time. Wolfgang Paul and his group in Bonn developed the three-dimensional Paul trap from a linear quadrupole mass filter, which can be regarded as a two-dimensional ion trap.[4] The Paul trap established the standard three-electrode structure for ion traps, consisting of a ring electrode and two endcap electrodes which generate the required electric field for trapping (see Section 1.3). Early work with the Paul trap included studies of the lifetime of ions held in the trap and spectroscopic measurements of hyperfine splittings in simple ions. A different research group developed a trap for charged macroscopic dust particles that operated on exactly the same principles as Paul's trap, but in a completely different parameter regime.[5] In this trap it was possible to observe the trajectories of the charged particles directly by using a camera.

The Penning trap, which makes use of a magnetic field for trapping, was reported a little later by a number of groups working independently, including Hans Dehmelt at the University of Washington,[6] and groups in Edinburgh,[7] Bonn[8] and Moscow.[9] It was named the Penning trap by Dehmelt in recognition of the fact that Frans Penning had (in 1936) reported the application of a magnetic field to an electrical discharge to prolong the lifetime of electrons due to the confining effect of the field.[10] The Penning trap was also used for a number of different types of studies, but in particular it was used for precision measurements of hyperfine splittings using techniques such as microwave-optical double resonance, and for measurements of the g-factor of the electron.

### 1.3. Principles of Ion Traps

All ion traps work by confining the motion of charged particles using electric and magnetic fields. The creation of a static three-dimensional potential well, which would be ideal for trapping, is forbidden by Earnshaw's theorem. Using static electric fields, it is only possible to create a saddle point in the potential. In the simplest case, a quadratic potential is created using three electrodes, as shown in Figure 1.1.

#### *1.3.1. The Penning trap*

The Penning trap[6] makes use of a static quadratic potential well along the axis joining the two endcap electrodes (the *z*-axis). This is created by putting a positive potential (*V*) on the endcaps relative to the ring (for positively charged particles). The resulting simple harmonic motion in the axial direction has an angular frequency of $\omega_z = (4qV/md^2)^{1/2}$, where *q* and *m* are the charge and mass of the ion respectively and *d* is a distance related to the separation of the endcaps ($2z_0$) and the diameter of the ring electrode ($2r_0$) by the equation $d^2 = r_0^2 + 2z_0^2$. Typical values of $d$ range from a few mm to a few cm, depending on the application.

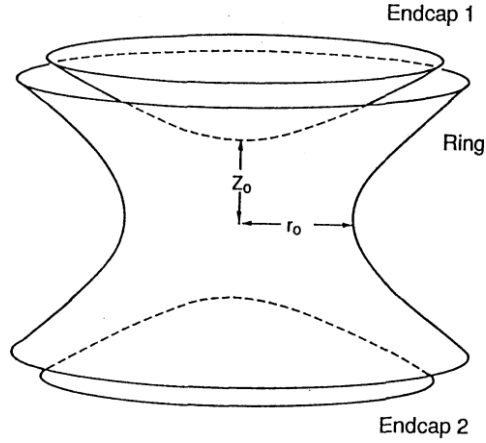

Fig. 1.1. Electrodes of an ion trap.

The potential in the radial plane also varies quadratically as a function of $x$ and $y$, but the potential has a maximum at $x=y=0$, making the radial motion unstable. This is overcome by applying a strong magnetic field along $z$. The magnetic field gives rise to a cyclotron-type motion of the ions at an angular frequency close to $\omega_c = qB/m$, which is the normal cyclotron frequency for a particle moving in a magnetic field of magnitude $B$. Its value is shifted down slightly by the presence of the radial electric field, which also gives rise to a second, slower, circular motion in the radial plane called the magnetron motion (at an angular frequency $\omega_m$). The modified cyclotron frequency and the magnetron frequency are given by

$$\omega'_c = \frac{\omega_c}{2} + \sqrt{\frac{\omega_c^2}{4} - \frac{\omega_z^2}{2}} \qquad (1.1)$$

$$\omega'_c = \frac{\omega_c}{2} - \sqrt{\frac{\omega_c^2}{4} - \frac{\omega_z^2}{2}}. \qquad (1.2)$$

The total energy associated with the magnetron motion is negative, due to its low frequency, combined with the negative radial potential energy, and this gives rise to several problems associated with the stability of ions in the trap. It also makes simultaneous cooling of all three degrees of freedom difficult.

Penning traps have been used for many different types of experiment, often using high magnetic fields in the range of 1–10 tesla from superconducting magnets. For trapped electrons this gives cyclotron frequencies up to tens of GHz. For protons the cyclotron frequencies are in the range of tens of MHz and for singly charged atomic ions they are typically of the order of 1 MHz. The axial frequency is restricted to values lower than $\omega_c/\sqrt{2}$ as otherwise the argument of the square root in Eq. 1.1 and Eq. 1.2 becomes negative.

Space charge limits the maximum number density of particles in the trap to values of the order of $10^8 - 10^9$ cm$^{-3}$ for atomic ions.

Penning traps also come in so-called open versions, referred to as cylindrical Penning traps or Penning–Malmberg traps, where there are no endcaps as such, but rather a number of co-axial cylindrical electrodes to generate the axially confining electric fields. These traps, while less suitable for precision measurements, offer increased versatility and facility of loading and unloading particles and are used in particular for studies of non-neutral plasmas, and for antimatter experiments.

### 1.3.2. *The radiofrequency (RF) Paul trap*

The RF trap[4] makes use of a dynamic trapping mechanism to overcome the instability associated with a static electric field. As was pointed out above, the saddle point from a positive potential applied to the endcaps gives stable motion in the axial direction and unstable motion in the radial direction. However, if the potential is reversed the radial motion is now stable and the axial motion unstable. The equations of motion for an oscillating potential (at angular frequency Ω) are examples of the well-known Mathieu equation. There are solutions that are stable in both the axial and radial directions, for a particular range of combinations of voltages and oscillation frequencies (at a given value of *q/m*). The final motion is then an oscillation in an effective potential that arises as a result of the driven motion. This so-called *pseudopotential* has a three-dimensional quadratic minimum at the centre of the trap and gives rise to simple harmonic motion in both the axial and radial directions. If theapplied potential is $V_{dc} + V_{ac}\cos(\Omega t)$ then we find (in a given parameter range[3]) the approximate oscillation frequencies

$$\omega_z = \frac{\Omega}{2}\sqrt{a_z + \tfrac{1}{2}q_z^2} \tag{1.3}$$

$$\omega_r = \frac{\Omega}{2}\sqrt{a_r + \tfrac{1}{2}q_r^2}, \tag{1.4}$$

where the stability parameters $a_z$ and $q_z$ are given by

$$a_z = -2a_r = -16qV_{dc}/m\Omega^2 d^2 \tag{1.5}$$

$$q_z = 2q_r = 8qV_{ac}/m\Omega^2 d^2. \tag{1.6}$$

Typically, the parameters are chosen such that the value of $q_z$ is in the region of 0.4 and $a_z$ is much less than $q_z$. Under these conditions the radial and axial frequencies are typically one-tenth of the applied frequency. The typical size of a trap designed for use with large clouds of ions is a few mm to a few cm.

Some experiments with small numbers of trapped atomic ions are carried out in miniature RF traps with electrode separations in the region of 1 mm. These traps are operated with applied frequencies of the order of a few MHz and a few hundred volts amplitude.

One consequence of the trapping mechanism in the RF trap is that there is always a small-amplitude driven motion at the applied frequency $\Omega$. This is referred to as the *micromotion*. (The only exception to this is if an ion is located exactly at the centre of the trap where the fields are zero.) The displacement from the centre can give rise to heating effects in some situations and also the micromotion can give rise to sidebands or broadening in the spectra of trapped ions.

One particularly important feature of the RF trap is its ability to reach the Lamb–Dicke regime.[11] This arises in optical spectroscopy, when the ion is better localized than a fraction of the wavelength of the interrogating laser light. As a consequence, the first order Doppler effect is eliminated and is replaced by a small number of sidebands separated by the oscillation frequency of the ion. This allows ultrahigh-resolution optical spectroscopy of forbidden transitions to be carried out. For optical transitions, this parameter regime can only be reached in a micromotion-free zone and is the indispensable condition for many experiments in frequency metrology or quantum information.

*1.3.3.  The linear RF trap*

A variant of the three-dimensional RF trap is the linear trap, where electrodes are again used to create an oscillating quadratic electrical potential.  However, in this case the electrodes are four parallel rods and the oscillating potential is applied between the rods, with opposite pairs being connected together. If we take the $z$-axis as the direction parallel to the rods, then the potential gives confinement in the $x$ and $y$ directions. In order to provide axial confinement as well, a direct current (DC) potential well is created, typically by applying positive voltages to two endcap electrodes at the ends of the rods.

The radial confinement strength is similar to that obtained in a conventional Paul trap, except that this type of trap is easier to make much smaller.  The oscillation frequencies can therefore be much higher, around 10 MHz in some cases.  The axial confinement depends only on the effective potential along the $z$-axis and is usually weaker than the radial one.

Linear RF traps have the advantage that they have a line where the micromotion is eliminated, rather than just a single point.  They are therefore often used in experiments where a small number of ions are required to be free of micromotion, for example in quantum information processing applications.  For this type of application they are often manufactured using microfabrication techniques, giving typical electrode separations in the region of a few hundred μm.

*1.3.4.  Low-energy storage rings*

An alternative way to store charged particles is to let them circulate in a storage ring. While storage rings are usually large machines intended for high-energy physics experiments, the last two decades or so have seen many machines built for low-energy applications.[12]

In a storage ring, or circular accelerator, particles are confined radially (i.e. perpendicular to their direction of motion) by alternating quadrupole fields that, depending on energy, may be either electrical (low energy) or magnetic (high energy)in origin. As a given quadrupole field will focus only in one plane, alternating focusing ensures confinement of the particles due to their forward motion. The similarity with the Paul trap is striking: in a storage ring it is the forward motion of the particles that ensures that the average transverse potentialis confining, and in the Paul trap it is the oscillating electric field that does the same. This setup with alternating focusing and defocusing elements is also referred to as an alternating gradient design.

Storage rings are beneficial for studies where, for example, a charge change is to be studied, allowing the charge-changed state to easily escape and be detected. By merging different beams they may also allow studies of very low-energy collisions, and with the somewhat recent reintroduction of electrostatic storage rings(over the last decade or so), storage and study of biomolecules has become possible.

For more details about the operation and applications of electrostatic storage rings, see the chapters by Papash in this volume.[13,14]

## 1.4. Creation, Cooling and Detection of Ions

Here we give a very short summary of important issues concerning the use of ion traps: how are ions loaded into the trap, how can they be cooled down for experiments and how are they detected?

### *1.4.1. Creation of ions*

There are two main approaches to the creation and loading of ions into an ion trap: either the ions need to be created inside the trap, or, if they are created elsewhere, they need to be captured by the trap.

In the first instance, ions can be created in two different ways. Initially, many experiments used electron bombardment to ionize atoms inside the trap volume (from a small atomic beam oven, for example). This is a very simple technique but is not at all selective, so any species that is present in the vacuum system may be ionized and trapped. However, the technique is very easy to implement.

More recently, photo-ionization has been used, particularly in experiments with single ions or small numbers of ions. This requires typically one laser source, tuned to a resonance transition in the neutral atom, and a second light source (also usually a laser) to excite the atom into the continuum by a further one or two steps, one of which may be resonant. Not only is this technique element-selective, it is also isotope-selective if the first laser has a narrow linewidth and can be tuned accurately to the resonance transition. In this way it has been possible to load even very low-abundance isotopes with high efficiency. The availability of a wide range of wavelengths from tunable diode lasers has enabled many research groups to move to photo-ionization as their preferred method of creating ions.

When ions are created inside the trap, they will automatically remain trapped (so long as the stability conditions are fulfilled). On the other hand, ions that have been produced externally (for example, from an ion source or accelerator) cannot simply be loaded into the trap because, although they will be attracted towards the centre of the trap, they will also have enough energy to exit the trap again. It is therefore usually necessary to find some way of turning the trap off, to allow the ions into the centre of the trap, and then turning the trap on again. So long as the ions do not have enough energy to escape, they will then remain trapped.

Alternatively, if there is a strong cooling mechanism present (e.g. buffer gas in the case of RF traps), ions that fly through the trap may be cooled enough while inside the trapping region to force them to remain trapped. In this case it is possible to accumulate ions over a period of time in order to fill the trap.

### 1.4.2. *Cooling of ions*

When ions are first loaded into a trap, it is likely that they will have large energies, comparable to the depth of the trap (that is, the energy required for an ion to escape). For example, if the ions are created throughout the trapping region, they will have a range of energies up to that which is required to exit the trap. Some of the higher energy particles will escape, and the rest will thermalize to a lower average energy, but this will still be higher than required for many experiments, especially where any spectroscopic technique is employed that is limited by the Doppler effect.

It is therefore frequently necessary to slow the ions down, i.e. to cool them. There are several useful outcomes from cooling ions: the energy of each particle is reduced, the ions are better localized at the centre of the trap, the ion density increases and the lifetime of the ions in the trap is increased. An increase in the number density may also allow plasma effects to be observed, in particular if the Debye length becomes much smaller than the dimensions of the ion cloud. Under these circumstances the ion cloud has a uniform density with a sharp boundary and is referred to as a non-neutral plasma.

Another effect that can arise at very low temperatures is crystallization of the ion cloud. This occurs if the thermal energy is much less than the Coulomb interaction energy between adjacent ions. The ions then take up fixed positions relative to each other and are said to form an *ion Coulomb crystal*. These crystals (the simplest of which is a string of ions along the axis of a trap) can be observed in images of cold ions in traps.

A number of techniques exist for reducing the energy of trapped ions. They have different advantages and are applicable to different types of experiments and ion species. More details can be found in the books referred to earlier[1–3] and in the chapter by Segal and Wunderlich in this volume.[15]

*1.4.2.1. Buffer-gas cooling*

The simplest technique is buffer-gas cooling. This allows the exchange of energy between the trapped particles and atoms (or molecules) of a low-pressure buffer gas, generally chosen to be light and inert (e.g. helium). The energy exchange takes place through thermal collisions and the final temperature achieved is limited to that of the apparatus, which is usually at room temperature. If the pressure is low enough, the stability characteristics of the trap are unaffected. The technique is not element-specific and is often used with clouds of ions in RFtraps. The choice of the buffer-gas species is important in order to ensure that collisions are elastic. The collisions with some molecular gases can be inelastic and change the internal energy state of the trapped ions, provoking quenching and state-mixing.

Buffer-gas cooling is complicated in Penning traps by the fact that the total energy of the magnetron motion is negative, so the thermalization process tends to lead to an increase in the magnetron radius rather than a decrease. However, it is used for cooling high-energy positrons as they are accumulated in a Penning trap. Buffer-gas cooling of both ions and positrons may be used in conjunction with other techniques that prevent the magnetron orbit size increasing (e.g. the rotating-wall technique, which uses an additional rotating electric field to prevent radial expansion of the plasma).

*1.4.2.2. Resistive cooling*

A technique that can be applied in the Penning trap is resistive cooling. In this arrangement a resistor is connected between the two endcap electrodes. As the ions move in the trap they induce image charges in the electrodes. These charges are continuously changing due to the motion of the ions, leading to a movement of real charges between the electrodes through the resistor. Energy is therefore dissipated in the resistor as heat, and this energy comes from the motional energy of the ions. It can be shown that this leads to an exponential decay of the energy of the axial motion. However, it is only the centre-of-mass motion that is cooled efficiently if a cloud consists of only one ion species. This is because the centre of charge then coincides with the centre of mass and this is stationary for all modes of oscillation except the centre-of-mass mode. Any internal energy of the ion cloud therefore does not generally couple strongly to the resistor. This can be improved, for example, by putting the resistor between one endcap and the ring rather than between the two endcaps so that the coupling is asymmetric and therefore becomes sensitive to internal modes of oscillation of the ion cloud. The coupling is also increased if there are ions of different species present, as the centres of charge and mass are now separate.

Again, the cooling can only reduce the ion temperature down to the temperature of the apparatus. This is because it is limited by Johnson noise in the resistor, which depends on temperature. One solution is to reduce the temperature of the apparatus (including the resistor) to around 4 K using liquid helium. In order to achieve a high rate of cooling, the effective value of the resistor can be increased by using a high quality factor (high-Q) tuned circuit.

Resistive cooling is effective for axial and cyclotron motions for atomic ions in the Penning trap but again is unsuitable for the magnetron motion. However, the magnetron motion can be coupled to either the cyclotron or axial motion by use of an additional oscillating potential, which is applied between appropriate electrodes at the sum frequency (the electrodes may need to be segmented for this purpose). If this is done, then energy is exchanged between the coupled motions and the resistive cooling becomes effective for all three motions. This is referred to as *sideband cooling* or *axialization* and is often used in experiments designed to measure the mass of trapped particles.

Note that for electrons in a trap the cyclotron motion automatically cools to the ground state in a short time via synchrotron radiation. This is only significant in the case of electrons and positrons because the cyclotron frequency is so high (up to tens of GHz at a field of a few tesla).

*1.4.2.3. Laser cooling*

The cooling technique that is able to reach the lowest temperatures is laser cooling. The simplest version, Doppler cooling, is very efficient and can typically take the temperature down to 1 mK. It involves irradiating the ion with laser light tuned close to a strong transition from the ground state of the ion. However, laser cooling is only applicable to a small number of species of ion: it has only been used for approximately ten different species, all of which are singly charged atomic ions (many of them alkali-like ions). Other species either have an energy level structure that is too complicated or do not have wavelengths that are accessible with tunable continuous-wave lasers.

Laser cooling works by transferring momentum from the laser beam to the ion and using the Doppler effect to ensure that the ion only absorbs light when it is moving towards the laser. The ion therefore slows down by a small amount each time a photon is absorbed from the beam. In principle this requires a two-level system so that after ions are excited by the laser they return by spontaneous emission to the initial state. However, in many species of interest there is a metastable state into which the ion can also decay. A second laser therefore needs to be provided to repump ions back out of this state into the cooling cycle, as otherwise no more cooling can take place. This is a disadvantage, but on the other hand the metastable states are often useful, for instance to use as a quantum bit ("qubit") for coherent processes in experiments in the area of quantum information processing, or for a clock transition for frequency standards.

Doppler laser cooling can be used in any sort of trap, but it is generally easier to apply in RF traps, as the strong magnetic field of the Penning trap leads to large Zeeman splittings and a requirement for many more laser frequencies. However, in some species without a metastable state, a single laser can still be used for Doppler cooling, taking advantage of optical pumping techniques.

In some experiments it is necessary to cool beyond the Doppler limit, even down to the ground state of the vibrational motion in the trap. This motion corresponds to a quantum mechanical simple harmonic oscillator and has states labeled by the vibrational quantum number $n$. In order to cool below the Doppler limit (which corresponds to the value of $n$ being typically of the order of 10) it is necessary to employ other techniques, the most common of which is sideband cooling. (This is not to be confused with the technique described above in the context of resistive cooling.) In this case the ion is cooled initially using Doppler cooling and then it is cooled further by exciting it on a narrow optical transition (generally a forbidden transition to a metastable state). If the transition is narrow enough, and the laser linewidth is small enough, the spectrum becomes a carrier with sidebands that arise from the vibrational motion in the trap. The sidebands are spaced at the vibrational frequency and their amplitudes are dependent on the degree of excitation of the motion. For appropriate trapping conditions, the ion will be close to the Lamb–Dicke regime and therefore there will only be a small number of sidebands present.

Assume that the initial vibrational quantum number is $n$. If the laser is tuned to the first sideband below the carrier frequency (the so-called red sideband) then when the ion is excited there is a loss of one quantum of vibrational excitation, so after excitation the ion is in the state $n - 1$. Spontaneous decay back to the ground state generally does not change the value of $n$. If this process of excitation and decay is repeated, eventually the ion will reach the ground state of the motion ($n = 0$), at which point the red sideband amplitude becomes zero as there is no state $n = -1$ available to the ion to be excited into.

Sideband cooling is used routinely in several laboratories for preparing individual ions (or several ions in a string) in the ground state of motion as a preparation for coherent operations such as quantum gates. These techniques are also used for trapped-ion atomic clocks (frequency standards).

*1.4.2.4. Sympathetic cooling*

Laser-cooling techniques are limited to a small number of ion species, as mentioned above. However, in some cases it is possible to use one laser-cooled species to cool a different species (present in the same trap) by a process of *sympathetic cooling*, mediated by the exchange of energy in thermal collisions between the ions. Sympathetic cooling has been used in both Penning and RF traps. It has been shown that in some cases a small number of laser-cooled ions can cool a much larger number of simultaneously trapped ions of other species by this process. Sometimes it results in a radial separation of the different species. Sympathetic cooling is also used for cooling of antiprotons by electrons held in the same trap. It has frequently been used with large clouds of ions but it also works when there are just two ions in the trap: this is the *quantum logic* approach, used in some recent frequency standards work. Here a laser-cooled ion is used to cool and also probe the electronic state of a second ion which has a highly stable and narrow optical transition, used as the clock transition.

*1.4.3.　Detection of ions*

Since atomic ions at low temperatures have very low energy, and often there are not many ions in a trap, the detection of trapped ions is often difficult. Here we list the commonly used methods for detecting the presence of ions in a trap. These methods fall into two groups: destructive detection and non-destructive detection. For more details on all aspects of the detection of ions in traps, see the chapterby Knoop in this volume.[16]

One method for detecting ions in a destructive manner is simply to release the ions from the trap and accelerate them towards a detector such as microchannel plate (MCP). This is a reliable and sensitive method of detection but, because it is destructive, it is necessary to reload the trap for each run of the experiment. This technique is often used in mass measurements for unstable isotopes created at accelerators.

One non-destructive technique is to use the image charges induced in the trap electrodes by the motion of ions in the trap – the same physical process used for resistive cooling. In this case the voltage across the resistor is monitored. When ions are present in the trap the induced current in the resistor gives rise to a voltage across it, which can be measured with a sensitive voltmeter. However, this is often too small to be seen directly, so an alternative is to make the electrodes part of a high-Q resonant circuit and to drive the motion using a weak excitation of this circuit. If ions are present, energy will be absorbed from the circuit, lowering its Q-value. This can be detected as a drop in voltage across the electrodes.

The most sensitive technique is to detect the laser fluorescence. This requires much the same energy level structure as laser cooling; indeed it is often the fluorescence from lasers used for laser cooling that is also employed to observe the ions. On resonance an ion may absorb and re-emit up to several million photons per second. With a solid angle for detection of 1% of $4\pi$ and an overall quantum efficiency of detection of up to 10%, this means that over 10 000 counts per second may be observed from a single laser-cooled ion. Indeed it is just possible to see an ion with the naked eye if the fluorescence is in the visible region of the spectrum (e.g. for $Ba^+$).

This signal may be recorded with a photon-counting photomultiplier. However, many experiments have used highly sensitive CCD cameras to record images of single ions or strings and crystals of laser-cooled ions and to study the properties of the structures that the ions form.

Laser fluorescence detection is also able in some cases to determine the electronic state of an ion. This needs an ion with a metastable state that is not addressed by the laser-cooling wavelengths. When the ion occupies this state, it is not able to absorb light, so no fluorescence is seen (until it decays from that state). Since otherwise thousands of photons would have been detected, this allows extremely efficient and sensitive detection of the electronic state of the ion. This was named the *electron shelving technique* by Dehmelt, who first introduced it, and it is now widely used in experiments involving coherent interactions between lasers and ions. It is routinely used to observe *quantum jumps* as an ion makes transitions into and out of a metastable state.

## 1.5. Applications of Ion Traps

Ion traps are now used in a large variety of applications throughout physics. Here we list some of the more important applications in order to give a flavour of the versatility of these devices.

*Precision measurements of fundamental quantities in atomic physics*

Due to the fact that ions in traps are so well isolated from the environment (e.g. collisions) and that they generally have low velocities, they are suitable for precision measurements of various sorts. An excellent example is the measurement of g-factors of electrons, positrons and atomic ions. In these experiments it is necessary to measure both the cyclotron frequency of the particle and the spin-flip frequency. The ratio of the two can be used to measure the g-factor to a high level of precision, without the need to know the strength of the magnetic field. These measurements make use of the fact that frequency is the physical quantity that can straightforwardly be measured to the highest accuracy (compared, for example, to length). The determination of the g-factor of the electron in a Penning trap has resulted in the most accurate value of a fundamental constant to date (with a precision of $3 \times 10^{-13}$).[17] Similar measurements with bound electrons in hydrogen-like ions have been used for tests of quantum electrodynamics (QED), as the g-factor is modified by the presence of the nucleus.

*Mass spectrometry*

Since the cyclotron frequency of an ion is inversely proportional to its mass, a measurement of the ratio of the cyclotron frequencies of two ions yields their mass ratio to a high level of precision. This approach has been used not only for the determination of the mass ratios of fundamental particles to the highest precision, but also for the measurements of long strings of stable and unstable isotopes at accelerator facilities such as CERN. These studies give vital information on nuclear binding energies used in the development of models of nuclear structure.[18]

*Storage of antimatter*

The extreme isolation of particles in a trap allows even exotic particles such as positrons and antiprotons to be stored for long periods of time. This has enabled studies of the creation of antihydrogen atoms to be carried out at CERN. For more details on the trapping of positrons in traps, see the chaptersby Surko in this volume.[19,20] The production of antihydrogen is discussed in the chapterby Madsen in this volume.[21]

*Non-neutral plasma studies*

A large cloud of charged particles in an ion trap constitutes a non-neutral plasma. So long as the Debye length is much less than the dimensions of the plasma, it will have a uniform density when it is in equilibrium. The confining potential of the trap is equivalent physically to a sea of particles of the opposite charge to those that are trapped. Non-neutral plasmas have been studied extensively, mainly in Penning traps, and techniques have been developed for manipulating the plasma density and shape as well as probing its modes of oscillation. Such experiments have been carried out with electrons, positrons and atomic ions. For more details of experiments on non-neutral plasmas in traps, see the chaptersby Anderegg in this volume.[22,23]

*Optical and microwave spectroscopy*

Since the early days of work with ion traps, they have been used as a way of confining a sample for study by spectroscopy. In this sense the detailed properties of the trap are not of interest as its main function is to keep the sample in the right place. One difficulty with this is that the maximum density of ions that can be held in a trap is limited by the Coulomb repulsion to typically $10^8$ ions per $cm^3$, which is similar to the number density of molecules in an ultrahigh vacuum system at a pressure of $10^{-9}$ mbar. This may result in signal levels in conventional spectroscopic investigations that are small compared to those carried out in low-pressure gases, for example. However, the extended interaction time available in an ion trap compensates for this disadvantage and so many important spectroscopic investigations have been carried out in the RF, microwave and optical regimes.

*Laser cooling*

As alluded to earlier, laser cooling enabled a wide range of experiments to be carried out in ion traps that were previously impossible, such as the observation of a single atomic particle at rest. Early experiments with laser cooling showed that single ions or small clouds of ions could be cooled to a temperature close to the Doppler limit fairly easily. This allowed the types of spectroscopic measurements referred to above to be taken to new levels of sensitivity and accuracy due in particular to the absence of the Doppler effect. Furthermore, developments in the theory of laser cooling showed that even lower temperatures could be reached using sideband cooling. In this way ions could be prepared in the ground state of the potential with high probability, opening the door to many new experiments. For more details on laser cooling, see the chapter by Segal and Wunderlich in this volume.[15]

*Microwave frequency standards*

One major application of laser-cooled ions in traps is the development of new frequency standards, both in the microwave region and the optical region of the spectrum. Penning traps are particularly well suited to microwave frequency standards as they can hold large numbers of laser-cooled ions in a single trap. The clock transition in the ions is driven by radiation from a stable oscillator, and is close to resonance with a microwave transition in the ions (which may be a hyperfine or Zeeman transition). The response of the ions to this radiation is detected by observation of the fluorescence from the ions, which depends on which state the ions are in. Using feedback to the oscillator that depends on the observed fluorescence, the oscillator can be stabilized to the ionic transition and can then be used as a highly stable reference, i.e. a frequency standard. Although some ion trap-based microwave clocks have been developed, they did not offer large benefits compared to well-developed atomic beam frequency standards such as the caesium clock.

*Optical frequency standards*

In the optical domain, ion traps have had a huge impact on frequency standards and are now performing at a much higher level of stability than the caesium clock. Most ion trap-based optical clocks use single ions that can be laser cooled on a strongly allowed optical transition. The ion also needs a narrow optical transition to a long-lived excited state that can be used both for sideband cooling and also as the clock transition. Much of the effort that has gone into ion trap research has been driven by the desire to develop new optical clocks. For this to succeed, it is necessary to be able to prepare a single ion in a trap, to cool it to the ground state of motion, to probe it with ultra-stable laser radiation, to detect its electronic state with high efficiency and to feed back to the laser in order to stabilize its frequency. These efforts have been so successful that a new definition of the second is likely to be based on an ion trap clock in due course. For more details on the development of ion trap-based atomic clocks, see the chapter by Margolis in this volume.[24]

*Ion Coulomb crystals*

When ions in traps are subject to strong laser cooling, eventually the thermal energy becomes much less than the energy of interaction of neighbouring ions via the Coulomb interaction. At some point the Coulomb interaction dominates so much that the ions fall into a crystal-like structure called an ion Coulomb crystal (ICC). As in a conventional crystal, this means that the ions take up fixed positions relative to each other. In very small crystals, and at the lowest temperatures, the ions do not exchange places. The simplest such structure is a linear string of ions, but it is also possible to see two-dimensional planar crystals and three-dimensional solid crystals. They have been observed in both Penning and RF traps with sizes ranging from a few ions up to many thousands of ions. This is a rich field of research in itself but also has applications, for example, in the study of chemical reaction rates between ions and neutral gases, where reaction rates may be measured by counting the number of ions in a crystal as a function of time.

*Quantum effects*

It is perhaps not surprising that the preparation of a single isolated atomic particle in the ground state of its motion allows interesting and novel quantum mechanical effects to be observed. Indeed trapped ions have been used in many such experiments, including the observation of quantum jumps in a single ion, the preparation and analysis of *Schrödinger cat states*, and the study of the quantum Zeno effect.

*Quantum information processing*

Perhaps the most exciting application of trapped ions has been in the field of quantum information processing. This field was initiated by Cirac and Zoller in 1995, who first proposed that a pair of laser-cooled trapped ions could be used to create a *quantum gate* – the quantum mechanical equivalent of a classical computer gate.[25] Practical realizations of their ideas followed and the field has grown rapidly ever since. Ion traps remain the system in which the most promising demonstrations of quantum information processing have been carried out. A large fraction of the current activity with trapped ions constitutes research in this area. It now includes quantum simulation as well as quantum computing applications. This work builds on many of the areas mentioned above, including laser cooling, spectroscopy and ICCs. For more details on all aspects of quantum information studies using ions in traps, see the chapterby Roos in this volume.[26] An alternative implementation of quantum information processing using ion traps is with trapped electrons as qubits. This is discussed in detail in the chaptersby Verdú in this volume.[27, 28]

*Quantum logic*

A recent development is the application of techniques from quantum information processing to frequency standards. One problem for frequency standards is to find an ion that has a suitable atomic structure for laser cooling but that also incorporates a narrow clock transition. A new solution is to separate these two functions so that one ion can be laser cooled but a different ion has the clock transition. When both are trapped at the same time, they form a Coulomb crystal of two ions and so the laser cooling is effective for both ions through the Coulomb interaction. Then a transition that takes place in the clock ion can be detected via a measurement technique based on the principles used in ion quantum gates, also mediated by the Coulomb interaction between the ions. This powerful technique has already been used to develop the most accurate ion-based optical clock to date, and another important application will be the measurement of possible variations of the fundamental constants with time.[24]

*Molecular Physics*

Electrostatic storage rings allow for the storage of a much larger mass range than conventional magnetic ones, and therefore several rings around the world are under construction or under commission for molecular physics experiments. Further, they are often small, and can therefore be cooled to cryogenic temperatures, an important feature for avoiding or studying blackbody radiation effects. An example from the negative ion world is the study of the lifetime of the negative $SF_6$ ion in the ELISA storage ring.[29] For more applications see the chapters by Papash in this volume.[13,14]

### 1.6. Conclusions and Outlook

Much has been achieved in the area of trapped ion research since ion traps were first invented in the late 1950s and early 1960s. These remarkable devices have found applications in many different areas of science and technology and have already led to the awarding of three Nobel Prizes. There is every sign that they will continue to find new applications in the coming years and will have even more impact in science and technology.

### Acknowledgments

This work was supported in part by the COST ACTION MP1001 "Ion Traps for Tomorrow's Applications", the CNRS, EPSRC, CERN and École de Physique des Houches.

### References

1. F. G. Major, V. N. Gheorghe and G. Werth, *Charged Particle Traps*(Springer,Berlin, Heidelberg, New York, 2004).
2. G. Werth, V. N. Gheorghe and F. J. Major,*Charged Particle Traps II: Applications*(Springer, Berlin, Heidelberg, New York, 2009).
3. P. K. Ghosh, *Ion Traps*(Oxford University Press, Oxford, 1995).
4. E. Fischer, Die dreidimensionale Stabilisierung von Ladungsträgern in einer Vierpolfeld, *Z. Phys.* **156,** 1–26 (1959).
5. R. F. Wuerker, H. Shelton and R. V. Langmuir, Electrodynamic containment of charged particles, *J. Appl. Phys.* **30,** 342–349 (1959).
6. H. G. Dehmelt, Radiofrequency spectroscopy of stored ions. I. Storage, *Adv. At.. Mol. Phys.* **3,** 53-72 (1967); H. G. Dehmelt, Radiofrequency spectroscopy of stored ions. II. Spectroscopy,*Adv. At.. Mol. Phys.***5**, 109–154 (1969).
7. J. Byrne and P. S. Farago, On the production of polarized electrons by spin-exchange collisions, *Proc. Phys. Soc.* **86,** 801–815 (1965).
8. G. Gräff and E. Klempt, *Z. Naturforschung***22a**, 1960–1962 (1967).

9. A. A. Sokolov and Yu. G. Pavlenko, *Optics and Spectroscopy* **22,** 1–3 (1967).
10. F. M. Penning, Introduction of an axial magnetic field in the discharge between two coaxial cylinders,*Physica***3**, 873–894 (1936).
11. R. H. Dicke, The effect of collisions upon the Doppler width of spectral lines, *Phys. Rev.* **89,** 472–473 (1953).
12. L. H. Andersen, O. Heber and D. Zajfman, Physics with electrostatic storage rings and traps, *J. Phys. B***37**, R57–R88 (2004).
13. A. I. Papash and C. P. Welsch, "Basics of beam dynamics and applications to electrostatic storage rings", in Knoop, M., Madsen, N. and Thompson, R. C.(eds), *Physics with Trapped Charged Particles*(Imperial College Press, London, 2013) pp. XXX–YYY.
14. A. I. Papash, A. V. Smirnov and C. P. Welsch, "Electrostatic storage rings at ultra-low energy range"in Knoop, M., Madsen, N. and Thompson, R. C. (eds), *Physics with Trapped Charged Particles*(Imperial College Press, London, 2013) pp. XXX–YYY.
15. D. M. Segal and Ch. Wunderlich,"Cooling techniques for trapped ions"in Knoop, M., Madsen, N. and Thompson, R. C. (eds), *Physics with Trapped Charged Particles*(Imperial College Press, London, 2013) pp. XXX–YYY.
16. M. Knoop, "Detection techniques for trapped ions" in Knoop, M., Madsen, N. and Thompson, R. C. (eds), *Physics with Trapped Charged Particles*(Imperial College Press, London, 2013) pp. XXX–YYY.
17. D. Hanneke, S. Fogwell and G. Gabrielse, New measurement of the electron magnetic dipole moment and the fine structure constant, *Phys. Rev. Lett.* **100**, 120801 (2008).
18. K. Blaum, Yu. N. Novikov and G. Werth, Penning traps as a versatile tool for precise experiments in fundamental physics, *Contemp. Phys.* **51,** 149–175 (2010).
19. C. M. Surko, "Accumulation, storage and manipulation of large numbers of positrons in traps I. – the basics" in Knoop, M., Madsen, N. and Thompson, R. C. (eds), *Physics with Trapped Charged Particles*(Imperial College Press, London, 2013) pp. XXX–YYY.
20. C. M. Surko, J. R. Danielson and T. R. Weber, "Accumulation, storage and manipulation of large numbers of positrons in traps II. – selected topics" in Knoop, M., Madsen, N. and Thompson, R. C. (eds), *Physics with Trapped Charged Particles*(Imperial College Press, London, 2013) pp. XXX–YYY.
21. N. Madsen, "Antihydrogen formation and trapping" in Knoop, M., Madsen, N. and Thompson, R. C. (eds), *Physics with Trapped Charged Particles*(Imperial College Press, London, 2013) pp. XXX–YYY.
22. F. Anderegg, "Waves in non-neutral plasma" in Knoop, M., Madsen, N. and Thompson, R. C. (eds), *Physics with Trapped Charged Particles*(Imperial College Press, London, 2013) pp. XXX–YYY.

23. F. Anderegg, "Internal transport in non-neutral plasmas" in Knoop, M., Madsen, N. and Thompson, R. C. (eds), *Physics with Trapped Charged Particles*(Imperial College Press, London, 2013) pp. XXX–YYY.
24. H. S. Margolis, "Optical atomic clocks in ion traps" in Knoop, M., Madsen, N. and Thompson, R. C. (eds), *Physics with Trapped Charged Particles*(Imperial College Press, London, 2013) pp. XXX–YYY.
25. J. I. Cirac and P. Zoller, Quantum computations with cold trapped ions, *Phys. Rev. Lett.***74,** 4091–4094 (1995).
26. C. F. Roos, "Quantum information processing with trapped ions" Knoop, M., Madsen, N. and Thompson, R. C.(eds), *Physics with Trapped Charged Particles*(Imperial College Press, London, 2013) pp. XXX–YYY.
27. J. Verdú, "Trapped electrons as electrical (quantum) circuits" in Knoop, M., Madsen, N. and Thompson, R. C. (eds), *Physics with Trapped Charged Particles*(Imperial College Press, London, 2013) pp. XXX–YYY.
28. J. Verdú, "Novel traps" in Knoop, M., Madsen, N. and Thompson, R. C. (eds), *Physics with Trapped Charged Particles*(Imperial College Press, London, 2013) pp. XXX–YYY.
29. J. Rajput, Measured lifetime of $SF_6^-$, *Phys. Rev. Lett.***100**, 153001 (2008).